\documentclass[times,twocolumn,final]{elsarticle}

\usepackage{medima}

\usepackage{amssymb}
\usepackage{latexsym}

\usepackage{url}
\usepackage{xcolor}

\usepackage{hyperref}

\usepackage{graphicx}
\usepackage{amsmath,amssymb}
\usepackage{xspace}
\usepackage{multirow,setspace,verbatim,amsfonts,amsmath,amsbsy,epsfig,url}
\usepackage{gensymb}
\usepackage{booktabs}
\usepackage{array,multirow}
\newcolumntype{P}[1]{>{\centering\arraybackslash}p{#1}}
\newcolumntype{M}[1]{>{\centering\arraybackslash}m{#1}}

\usepackage{algorithm}
\usepackage[noend]{algpseudocode}
\usepackage{tikz}
\usepackage{mathrsfs} 
\usepackage[algo2e]{algorithm2e} 
\usepackage{array}
\usepackage{color}
\usepackage{tabularx}
\usepackage{makecell}
\usepackage{bm}

\newcommand{\Fb}{\boldsymbol{F}}
\newcommand{\Zb}{\boldsymbol{Z}}
\newcommand{\Mb}{\boldsymbol{M}}
\newcommand{\Yb}{\boldsymbol{Y}}
\newcommand{\Sb}{\boldsymbol{S}}
\newcommand{\fb}{\boldsymbol{f}}
\newcommand{\xb}{\boldsymbol{x}}
\newcommand{\zb}{\boldsymbol{z}}
\newcommand{\Rd}{\mathbb{R}}
\newcommand{\Gcb}{\boldsymbol{\mathcal{G}}}
\newcommand{\Ncb}{\boldsymbol{\mathcal{N}}}

\newcommand{\Tcb}{\boldsymbol{\mathcal{T}}}

\definecolor{newcolor}{rgb}{.8,.349,.1}

\begin{document}

\verso{S. Park, G. Kim and JC Ye \textit{et~al.}}

\begin{frontmatter}

\title{Vision Transformer  using  Low-level Chest X-ray Feature Corpus for COVID-19 Diagnosis and Severity Quantification}%

\author[1]{Sangjoon Park\fnref{fn1}}
\fntext[fn1]{Sangjoon Park and Gwanghyun Kim are co-first authors.}
\author[1]{Gwanghyun Kim\fnref{fn1}}
\author[1]{Yujin Oh}
\author[2]{Joon Beom Seo}
\author[2]{Sang Min Lee}
\author[3]{Jin Hwan Kim}
\author[4]{Sungjun Moon}
\author[5]{Jae-Kwang Lim}
\author[1]{Jong Chul Ye\corref{cor1}}
\cortext[cor1]{Corresponding author: 
  Tel.: +82-42-350-4320;  
  fax: +82-42-350-4310;}
\ead{jong.ye@kaist.ac.kr}

\address[1]{Korea Advanced Institute of Science and Technology (KAIST), Daejeon, South Korea}
\address[2]{Asan Medical Center, University of Ulsan College of Medicine, Seoul, South Korea}
\address[3]{College of Medicine, Chungnam National Univerity, Daejeon, South Korea}
\address[4]{College of Medicine, Yeungnam University, Daegu, South Korea}
\address[5]{School of Medicine, Kyungpook National University, Daegu, South Korea}

\received{}
\finalform{}
\accepted{}
\availableonline{}
\communicated{}

\begin{abstract}

Developing a robust algorithm to diagnose and quantify the severity of COVID-19 using Chest X-ray (CXR) requires a large number of well-curated COVID-19 datasets, which is difficult to collect under the global COVID-19 pandemic. On the other hand,  CXR data with other findings are abundant. This situation is ideally suited for the Vision Transformer (ViT) architecture, where a lot of unlabeled data can be used through structural modeling by the self-attention mechanism. However, the use of  existing ViT is not optimal, since feature embedding through direct patch flattening or ResNet backbone in the standard ViT is not intended for CXR. To address this problem, here we propose a novel Vision Transformer that utilizes low-level CXR feature corpus obtained from a backbone network that extracts common CXR findings. Specifically, the backbone network is first trained with large public datasets to detect common abnormal findings such as consolidation, opacity, edema, etc. Then, the embedded features from the backbone network are used as corpora for a Transformer model for the diagnosis and the severity quantification of COVID-19. We evaluate our model on various external test datasets from totally different institutions to evaluate the generalization capability. The experimental results confirm that our model can achieve the state-of-the-art performance in both diagnosis and severity quantification tasks with superior generalization capability, which are sine qua non of widespread deployment.

\end{abstract}


\end{frontmatter}


\begin{figure*}[!t]
\centering
\includegraphics[width=0.9\textwidth]{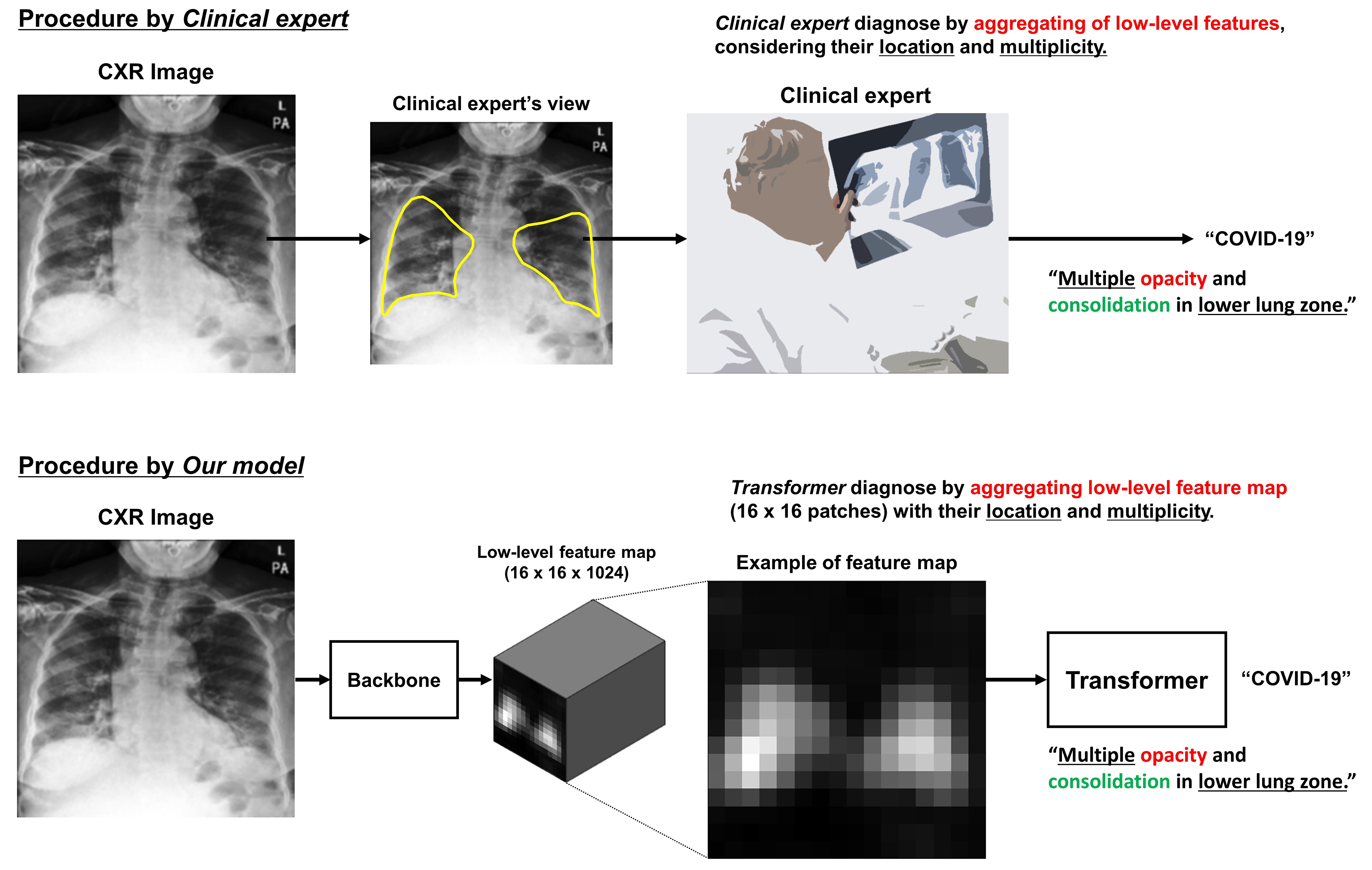}
\caption{Analogy between the diagnosis by a clinical expert and by our method.} \label{fig1}
\end{figure*}

\section{Introduction}
\label{sec1}
The novel coronavirus disease 2019 (COVID-19) caused by severe acute respiratory syndrome coronavirus 2 (SARS-CoV-2) has emerged as one of the deadliest virus of the century, resulting in about 137 million people infected with over 2.9 million death worldwide as of April 2021. In the light of the unprecedented pandemic of COVID-19, public health systems have faced many challenges, including scarce medical resources, which are pushing healthcare providers to face the threat of infection
 \citep{ng2020covid}. Considering its ominously contagious nature, the early screening of COVID-19 infection becoming increasingly important to avert further spread of disease and thereby reduce the burden on saturated health care system.

Currently, the real-time polymerase chain reaction (RT-PCR) is considered as gold standard in diagnosis of COVID-19 for its high sensitivity and specificity \citep{tahamtan2020real}, but it takes several hours and even days depending on regions to get the  exam results due to overstressed laboratories. Since the majority of patients with confirmed COVID-19 present positive radiological findings, the radiologic examinations can be useful for rapid screening of disease \citep{shi2020radiological}. Although computed tomography (CT) scan has excellent sensitivity and specificity for COVID-19 diagnosis \citep{bernheim2020chest}, the  use of CT is a major burden because of its high cost and potential for cross-contamination in the radiology suite.
Therefore, CXR holds many practical advantages as primary screening tool in the pandemic situation. 
In addition, CXR is useful for follow-up, which should be inexpensive and low in radiation exposure, to assess response to treatment.

Consequently, many studies have reported early application of CXR deep learning for  diagnosis \citep{wang2020covid, hemdan2020covidx, narin2020automatic, oh2020deep} or severity quantification of COVID-19 \citep{cohen2020predicting, sig2020covid, zhu2020deep, wong2020covidnet}, but they suffered from ineradicable drawbacks of poor generalization capability stemming from the scanty labelled COVID-19 data \citep{hu2020challenges, zech2018variable,roberts2021common}. The stable generalization performance on unseen data is indispensable for widespread adoption of the system \citep{roberts2021common}. 

One of the most commonly used measures to solve this problem is to build a robust model with innumerable training data \citep{chen2020more}, but it is difficult to construct large-scale dataset with labeled COVID-19 cases under the current pandemic situation. As a result, several methods have been proposed to mitigate the problem by transfer learning \citep{apostolopoulos2020covid}, weakly supervised learning \citep{zheng2020deep,wang2020weakly}, and anomaly detection \citep{zhang2020covid}, but their performances are still suboptimal.

The previous studies mostly utilize convolutional neural network (CNN) models, which was not specially designed for manifestations of COVID-19 which can be characterized by bilateral involvement, peripheral and lower zone dominance of ground glass opacities and patchy consolidations \citep{cozzi2020chest}. Although CNN architecture has shown to be superb in many vision tasks, it may not be optimal for problems requiring high-level CXR disease classification, where the global characteristics like multiplicity, distribution, and patterns have to be considered. This is due to the 
the intrinsic locality of pixel dependencies in the convolution operation.

 To overcome the similar limitation of CNN in computer vision problems that requires the integration of global relationshop between pixels, Vision Transformer (ViT) equipped with the Transformer architecture  \citep{vaswani2017attention} was proposed to model long-range dependency among pixels through the self-attention mechanism, showing the state-of-the-art (SOTA) performance in image classification task \citep{dosovitskiy2020image}. 
Since the Transformer
   was originally invented for natural language processing (NLP) in order  to attend different positions of the input sequence within a corpus and compute a representation of that sequence,
the choice of an appropriate corpus is the prerequisite for  the Transformer design.
%
 
In the original paper  \citep{dosovitskiy2020image}, two ViT models were suggested utilizing either direct pixel-patch embedding or feature embedding by ResNet backbone as corpora for Transformer. 
A problem occurs here, however, that neither the direct pixel-patch embedding nor feature embedding from ResNet may not be the optimal input embedding for CXR diagnosis of COVID-19. Fortunately, several large-scale CXR data sets are  constructed before COVID-19 pandemic and publicly available. For example, CheXpert \citep{irvin2019chexpert}, a large dataset that contain over 220,000 CXR images, provides labeled common low-level CXR findings (e.g. consolidation, opacity, edema, etc.), which is also useful for diagnosis of infectious disease. Moreover, an advanced CNN architecture has been suggested using the same dataset \citep{ye2020weakly}, which uses probabilistic class activation map (PCAM) pooling to leverage the class activation map to enhance the localization ability as well as classification performance. To take the maximum advantage of both the dataset and the network architecture for COVID-19, here we propose a novel ViT architecture which utilizes this advanced CNN architecture as a feature extractor for low-level CXR feature corpus, upon which Transformer is trained for downstream tasks of diagnosis  by utilizing the self-attention mechanism in Transformer.

It is worth to mention that our network basically identical to the text classification task with Transformer architecture, where the Transformer not only add up the meaning but also consider the location and relationship of words to make classification in sentence-level. Moreover, our method emulates the clinical experts who determine the final diagnosis of CXR (e.g. normal, bacterial pneumonia, COVID-19 infection, etc.) by comprehensively considering the low-level features with their pattern, multiplicity, location and distribution (e.g. \emph{Multiple} opacities and patch consolidations exist with \emph{lower lung zone dominance}: high probability for COVID-19) as illustrated in Fig. \ref{fig1}.

Another important contribution of this paper  is to show that our ViT framework can be also used for COVID-19 severity quantification and localization, enabling the serial follow-up of severity and thereby assisting the treatment decision of clinicians \citep{cohen2020covid}. The severity of COVID-19 can be determined by quantifying the extent of COVID-19 involvement. 
Recently,  array-based simple severity annotations where 1 or 0 is assigned to each 6 subdivisions of lungs are proposed by \citet{toussie2020clinical}, and we are interested in utilizing this weak labeling approach for severity quantification.
As the Transformer output already incorporates the long-range relationship between regions through self-attention, we use this Transformer output  to design a light-weighted network that  can accurately
quantify and localize the COVID-19 extents from weak labels.
Specifically, we adopt  the region of interest (ROI) max-pooling of the output Transformer feature to bridge the severity map and simple array. 
Consequently, in addition to the global severity score from 0 to 6, our model can create an intuitive severity level map where each pixel value explicitly means the likelihood of the presence of a COVID-19 lesion using 
the weak array-based labels.
In summary,  our main contributions are as follows.
\begin{itemize}
\item A novel ViT model for COVID-19 is proposed by leveraging the low-level CXR feature corpus that contain the representations for common CXR findings with pre-built large-scale dataset.
\item We have not limited our model to diagnosis, but expanded our model to quantify severity to provide clinicians with clinical guidelines for making treatment decisions.
\item We experimentally demonstrated that our method outperforms other Transformer-based models as well as CNN-based models especially in terms of the generalization on unseen data.
\end{itemize}

The remainder of this paper is organized as follow. Section \ref{sec2} summarizes the related works. Section \ref{sec3} and Section \ref{sec4} describes the proposed framework and datasets, respectively. Experimental results are presented in Section \ref{sec5}. Finally, we conclude this work in Section \ref{sec6}.

\begin{figure*}[!t]
\centering
\includegraphics[width=0.95\textwidth]{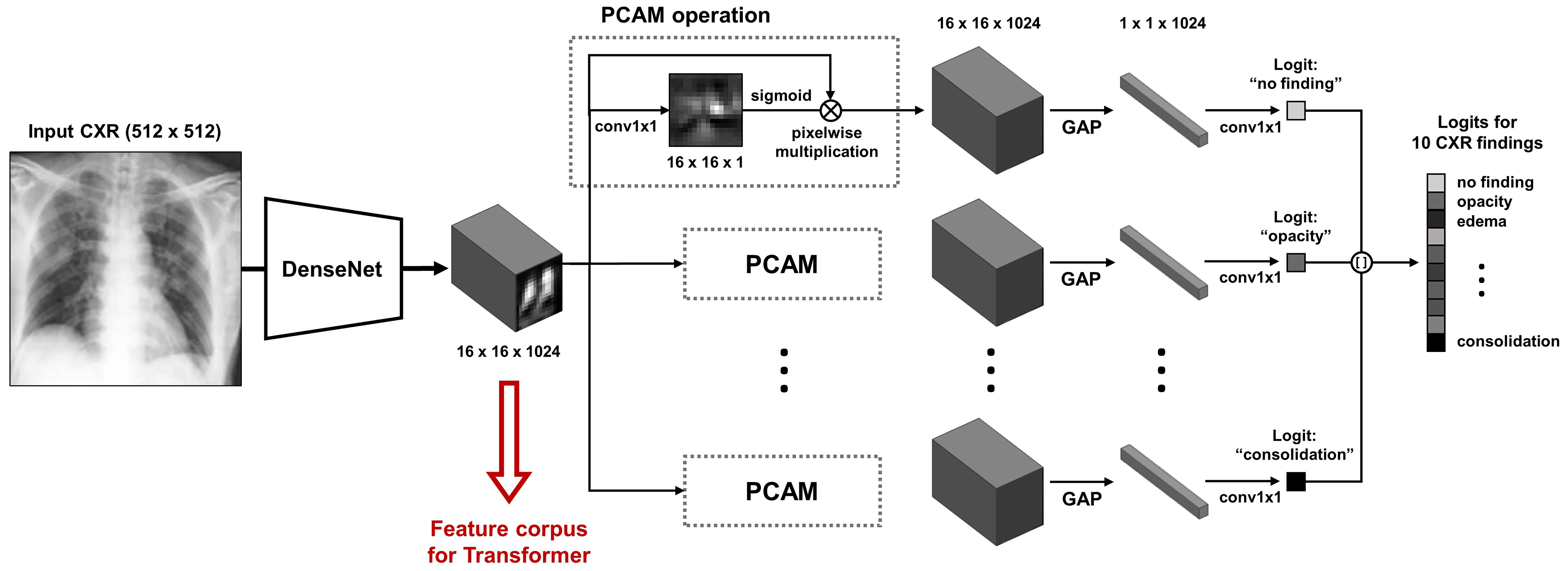}
\caption{Backbone network to extract low-level CXR feature corpus.} 
\label{fig2}
\end{figure*}

\section{Related work}
\label{sec2}
\subsection{Vision Transformer}
Transformer \citep{vaswani2017attention}, which was originally invented for NLP, is a deep neural network based on self-attention mechanism that facilitates appreciably large receptive fields. After demonstrating its astounding performance,
not only has Transformer become a de facto standard practice in NLP, but it has also motivated the computer vision community to explore its applications in computer vision by taking advantage of the long-range dependency between pixels \citep{khan2021transformers}. 

The ViT was the first major attempt to apply a pure Transformer directly to image, suggesting that it can completely replace the standard convolution operations by attaining SOTA performance. However, the experimental results showed that training vanilla ViT model requires a huge computational cost. 
Therefore, the authors also suggested hybrid architecture by conjugating CNN backbone (e.g. ResNet) to Transformer. With the feature extracted by ResNet, the Transformer can mainly focus on modeling the global attention. The experimental results suggest that it was able to achieve higher performance with the hybrid approach with relatively small amount of computations. 

After the introduction of ViT, the application of Transformer in computer vision has become an active area of investigation, resulting in many variant models of ViT showing SOTA performance in a variety of vision tasks including object detection \citep{zhu2020deformable}, classification \citep{dosovitskiy2020image, chen2020generative}, segmentation \citep{zheng2020rethinking}, and so on.

\subsection{Probabilistic Class Activation Map Pooling}
Class activation map (CAM) is a sort of class-specific saliency map obtained by quantifying the contribution of particular area of an image to the prediction of network. The most useful aspect of CAM is that it enables the localization of the important area only with weak labels, namely image-level supervision. Despite its excellent localization ability, most of previous works utilized CAM to generate heatmaps for lesion localization and visualization during inference. To leverage the localization ability of CAM to enhance the performance of network itself, one recent study utilized the CAM during training in CXR classification and localization tasks \citep{ye2020weakly}. They devised a novel global pooling operation that explicitly leverages the CAM in a probabilistic manner and is known as Probabilistic-CAM (PCAM) pooling. Different from standard approaches that use CAM for direct localization, they bound it with additional fully-connected layer and sigmoid function to get probabilities for each CXR findings. Then, the normalized attention weights were obtained from these output probability to make weighted feature maps containing more useful representation for each class. They showed that PCAM pooling operation can enhance both localization and diagnostic performance of  the model and achieved  first place in the 2019  CheXpert Challenge.

\subsection{COVID-19 Severity Quantification}
To build an automated algorithm for severity quantification, pixel-level annotation such as lesion segmentation label can offer a plentiful information. However, this type of labelling methods are labor-intensive 
and collecting large data with this pixel-level annotated label is not feasible under the global pandemic of COVID-19. To alleviate the problem,  simplified severity annotation methods, such as score-based and array-based methods, have been proposed. For example, \citet{cohen2020predicting} suggested a geographic extent score and a lung opacity score based on a rating system of lung oedema proposed by \citet{warren2018severity}. A geographic extent score assigns scores that range from 0 to 4, while lung opacity score assigns values of 0 to 3 based on the severity of involvement in each lung area. \citet{borghesi2020covid} designed Brixia score, another array-type severity labeling method, dividing lung with anatomic landmarks and assign score of 0-3 to each subdivision. Similarly, \citet{toussie2020clinical} suggested an array-based severity score for COVID-19. After dividing both lungs into six divisions, each area is assigned a value of 0 or 1, depending on the presence of COVID-19 involvement, which adds up to an overall severity of 0 to 6. We adopted the array-based annotation method suggested by \citet{toussie2020clinical} for severity quantification of COVID-19.

\begin{figure*}[!h]
\centering
\includegraphics[width=0.9\textwidth]{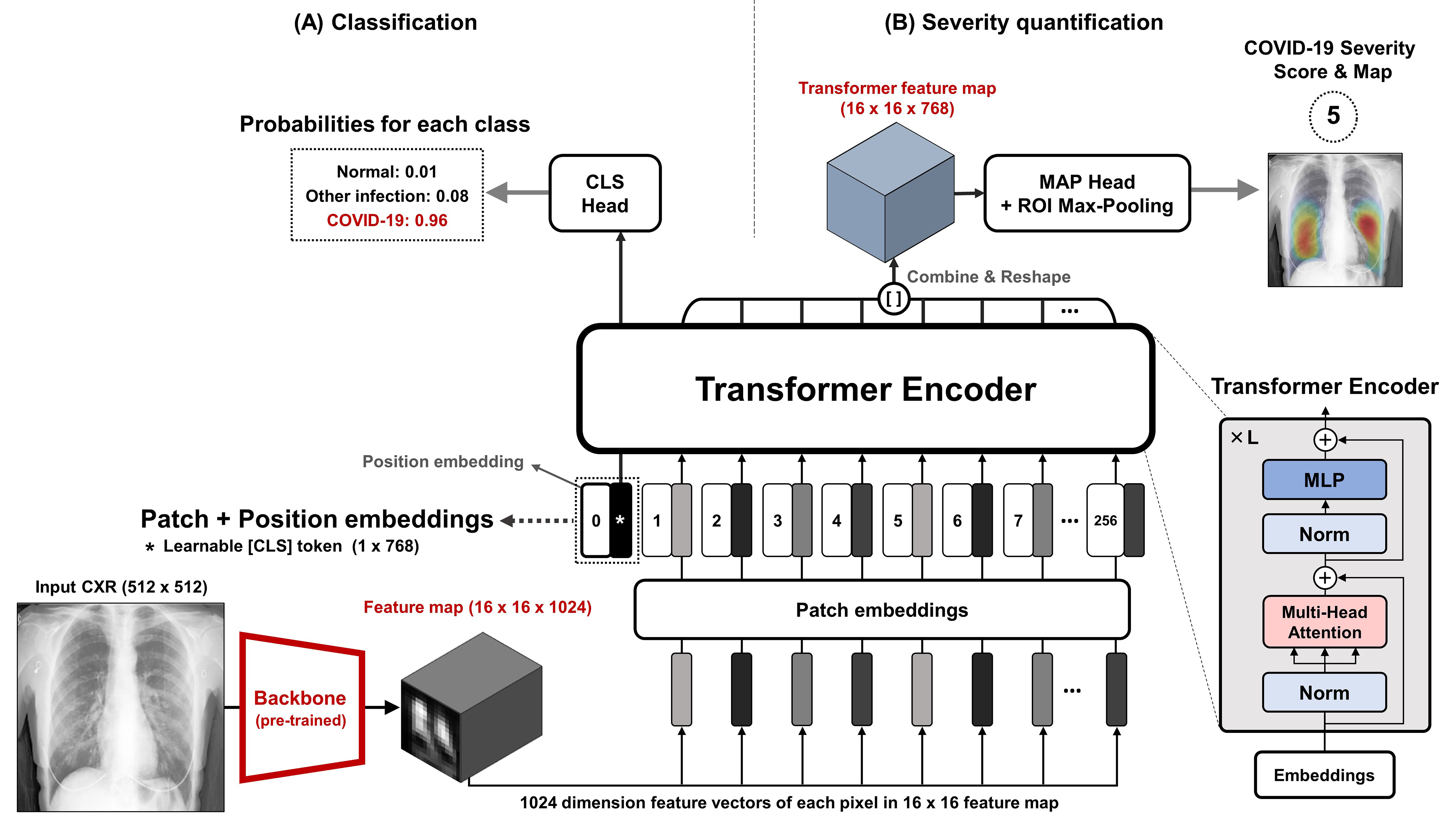}
\caption{Proposed framework and Vision Transformer model for diagnosis and severity quantification of COVID-19 on CXR for (A) disease classification, and (B) severity quantification.} 
\label{fig3}
\end{figure*}

\section{The Proposed Framework}
\label{sec3}
One of the novel contributions  of our approach is to show that we can maximize the performance of the Transformer model by using the low-level CXR corpus that comes from the backbone network trained with a large well-curated public record to produce common CXR findings.
 As the backbone network is trained with large number of data, the network is less prone to overfitting and thereby the generalization capability can be improved.

\subsection{Pre-training Backbone Network for Low-level Feature Corpus}
As a backbone network to extract low-level feature, we used the modified version of network proposed by \citet{ye2020weakly}. Firstly, the backbone network was pre-trained to classify 10 common low-level findings with a large public dataset. As depicted in Fig. \ref{fig2}, feature maps in each layer can be the candidates for utilizable feature embedding for the subsequent Transformer, and we experimentally found that the common embedding before the PCAM operation comprises of  most useful information. Nevertheless, care should be exercised since the PCAM operation for specific low-level CXR findings (e.g. lung opacity, consolidation, etc.) turns out to be crucial to achieve the optimal embedding at intermediate level, as  PCAM aligns these features to obtain the better classification results. More detailed experimental results about the level of feature map will be provided using ablation studies in Section \ref{sec5}.6.2.

\subsection{Vision Transformer for Classification}
The overall framework and the architecture of our ViT model is provided in Fig. \ref{fig3}. 
Specifically,  for a given $H\times W$ size input image $\boldsymbol{x} \in \mathbb{R}^{{H}\times{W}}$,
 the backbone network $\Gcb$  generates  $H'\times W'$ size feature maps $\Fb$:
\begin{flalign}
&& \Fb&= \boldsymbol{\mathcal{G}}(\boldsymbol{x}) &
\end{flalign}
Here,  the feature tensor  $\Fb \in \mathbb{R}^{{H'}\times{W'}\times{C'}}$ is defined as
\begin{flalign}
&&\Fb &= \begin{bmatrix}\fb_{1} &\fb_{2} &\cdots & \fb_{{H'}\times{W'}}\end{bmatrix} &
\end{flalign}
where
 $\fb_n \in \mathbb{R}^{{C'}}$   denotes a $C$-dimensional  embedded representation of low-level features at the $n$-th encoded block.
 These feature vectors  are used 
 to construct the low-level CXR feature corpora for Transformer.

Then, similar to Bidirectional Encoder Representations from Transformers  (BERT) \citep{devlin2018bert}, our Vision Transformer uses Transformer encoder layers to the input embedding. Specifically, since the Transformer encoder utilizes constant latent vector of dimension $D$, the extracted $C'$ dimension feature $\fb_n \in \mathbb{R}^{{C'}}$ is first projected to a $D$ dimension feature $\tilde \fb_n \in \mathbb{R}^{{D}}$ using $1$ $\times$ $1$ convolution kernel.
 We then prepended learnable \texttt{[class]} token embedding vector $\fb_\texttt{cls} \in \mathbb{R}^{{D}}$ to projected feature tensor.
This leads to the following composite projected feature tensor:
\begin{align}
\tilde\Fb &= \begin{bmatrix}\fb_\texttt{cls} & \tilde\fb_{1} &\tilde\fb_{2} &\cdots & \tilde\fb_{{H'}\times{W'}}\end{bmatrix}
\end{align}
 A positional embedding $\textbf{E}_{pos}$ that has the same shape to the projected feature tensor $\tilde\Fb$ is then added to encode a notion of the sequential order:
 \begin{align*}
 \Zb^{(0)} 
 &=\tilde\Fb + \textbf{E}_{pos}
 \end{align*}
This is then used as an input to a Transformer composed of $L$ successive encoder layers:
\begin{align}
\Zb^{(l)} = \Tcb^{(l)}\left(\Zb^{(l-1)} \right),\quad l=1,\cdots,L
\end{align}
where $\Zb^{(l)}=\begin{bmatrix}\zb_0^{(l)}&\zb_1^{(l)}& \cdots & \zb_{H'\times W'}^{(l)} \end{bmatrix}$
and $\Tcb^{(l)}$ denotes the $l$-th encoder layer.
The encoder layers used in our model are the same as standard Transformer which consists of repeated layers of multihead self-attention (MSA), multilayer perceptron (MLP), layer normalization (LN), and residual connections in each block, as shown in Fig.~\ref{fig3}.
Then, the first column $\zb_0^{(L)}$ of $\Zb^{(L)}$ represents
the Transformer attended feature vector with respect to
%
%
%
%
%
 the \texttt{[class]} token.
 Therefore, by simply adding a linear classifier as the classification head,
 we can obtain the diagnosis result of the input CXR image $\xb$ (see Fig.~\ref{fig3}(A)).
 

\begin{figure*}[!t]
\centering
\includegraphics[width=0.85\textwidth]{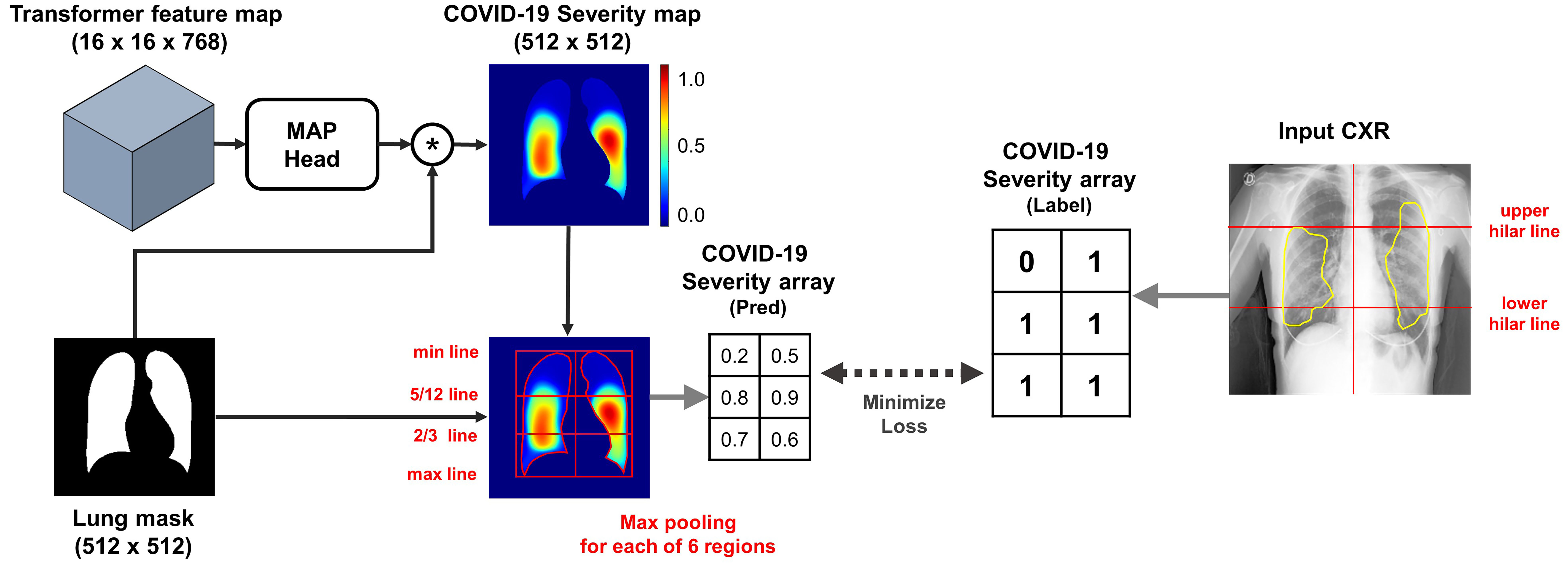}
\caption{Procedure of severity prediction and labeling. (A) Map head and ROI max-pooling of proposed framework. (B) Our severity annotation method for severity quantification on CXRs.} 
\label{fig4}
\end{figure*}

For the interpretability of the classification model,
we adopted a visualization method of saliency map tailored for ViT suggested by \citep{chefer2020transformer}, which computes relevancy for Transformer network.  Specifically,
unlike the traditional approaches of gradient propagation methods \citep{selvaraju2017grad, smilkov2017smoothgrad, srinivas2019full} or attribution propagation methods \citep{bach2015pixel, gu2018understanding}, which rely on the heuristic propagation along attention graph or the obtained attention maps, the method  in \citet{chefer2020transformer} calculate the local relevance with deep Taylor decomposition, which is then propagated throughout the layers. This relevance propagation method is especially useful for models based on Transformer architecture, as it overcomes the problem of self-attention operations and skip connections.

\subsection{Vision Transformer for Severity Quantification}
In the classification task, only one Transformer output at the \texttt{[class]} token position is used.
However, the rest of the Transformer output also produces feature embedding at each block position by taking into account of long-range relations
between the blocks. Therefore, we conjecture that  this information is useful
 for the severity quantification, as severity is determined by both local and global manifestation of the disease.
Accordingly, as shown in Fig. \ref{fig3}, these outputs are combined by additional lightweighted network to produce the COVID-19 severity map.

Specifically, as shown in Figs. \ref{fig3}(b) and  \ref{fig4},
 we first extract the Transformer output $\Zb^{(L)}$ except the \texttt{[class]} token position:
\begin{align}
\Zb_{res}   &=  \begin{bmatrix}\zb_1^{(L)} & \cdots & \zb_{H'\times W'}^{(L)}\end{bmatrix}
\end{align}
which is used as an input to the map head network $\Ncb$
\begin{align}
\Sb=\Ncb(\Zb_{res})
\end{align}
Then, the network output is multiplied pixel-wise with the segmentation mask $\Mb$, after which
ROI max-pooling (RMP) is applied to generate the severity mask $\Yb_{sev}\in \Rd^{3\times 2}$:
\begin{align}
\Yb_{sev} =  \texttt{RMP}\left(\Sb\otimes \Mb\right)
\end{align}
where $\otimes$ denotes the Hadamard product. In detail, the lung was divided into a total of six subdivisions, by dividing the right and left lungs into three subdivision (upper, middle, lower zone) with 5/12 and 2/3 line. Next, the largest values within each six subdivision were assigned as predicted values of severity array. Then, the map head network is trained by minimizing error of the estimated severity array with respect to the weakly annotated severity label as in Fig.~\ref{fig4}.


To generate the lung segmentation mask, we used method introduced by \citet{oh2021unifying}. 
In contrast to the existing approaches that are prone to under-segmentation for the severely infected lung with large consolidations,
this novel approach enables the accurate segmentation of abnormal lung as well as normal lung area by learning common features
using
 a single generator with AdaIN layers. 
 Since a single generator is used for all these tasks  by simply changing the AdaIN codes, the generator can synergistically learn the common features
 to improve segmentation performance for abnormal CXR data.


\begin{table*}[!t]
  \caption{Summary of dataset resources and disease classes for PA view CXRs.}
  \label{table:dataset_PA}
  \centering
  \resizebox{0.90\textwidth}{!}
  {%
  \begin{tabular}{M{2cm}|M{1.5cm}|M{1.5cm}|M{1.5cm}|M{1.5cm}|M{1.5cm}|M{1.5cm}|M{1.5cm}|M{1.5cm}}
  \Xhline{1.2pt}
  \multirow{2}{*} {\textbf{PA view}} & \multirow{2}{*} {\textbf{Total}} & \multicolumn{3}{c|}{\textbf{External test}} & \multicolumn{4}{c}{\textbf{Training and Validation}}\\ 
  \cline{3-9}
  & & {CNUH} & {YNU}& {KNUH}  & {AMC} & {NIH} & {Brixia} & {BIMCV} \\
  \cline{2-3} \cline{6-7}
  \hline
   \multicolumn{1}{l|}{Normal} & 13,649 & 320 & 300 & 400 & 8,861 & 3,768 & {-} & {-}\\
  \multicolumn{1}{l|}{Other infection} & 1,468 & 39 & 144 & 308 & 977 & {-} & {-} & {-} \\
   \multicolumn{1}{l|}{{COVID-19}}  & 2,431 & 6 & 8 & 80 & {-} & {-} & 1,929 & 408\\
   \cline{1-9}
   \multicolumn{1}{l|}{{Total images}}  & 17,548 & 365 & 452 & 788 & 9,838 & 3,768 & 1,929 & 408\\
   \Xhline{1.2pt}
  \end{tabular}}
\end{table*}

\begin{table*}[!t]
  \caption{Summary of dataset resources and disease classes for AP view CXRs.}
  \label{table:dataset_AP}
  \centering
  \resizebox{0.98\textwidth}{!}
  {%
  \begin{tabular}{M{2.0cm}|M{1.2cm}|M{2.0cm}|M{1.2cm}|M{1.2cm}|M{1.2cm}|M{1.2cm}|M{1.2cm}|M{1.2cm}|M{1.2cm}|M{1.2cm}}
  \Xhline{1.2pt}
  \multirow{2}{*} {\textbf{AP view}} & \multirow{2}{*} {\textbf{Total}} & \textbf{External test} & \multicolumn{8}{c}{\textbf{Training and Validation}}\\ 
  \cline{3-11}
  & & {CNUH} & {YNU}& {KNUH}  & {AMC} & {NIH} & {CheXpert} & {PADchest} & {Brixia} & {BIMCV} \\
  \cline{2-3} \cline{6-11}
  \hline
   \multicolumn{1}{l|}{Normal} & 14,507 & 97 & {-} & {-} & 117 & 3,390 & 9,500 & 1,310 & {-} & 93\\
  \multicolumn{1}{l|}{Other infection} & 204 & 19 & 76 & 92 & 17 & {-} & {-} & {-} & {-} & {-} \\
   \multicolumn{1}{l|}{{COVID-19}}  & 3,334 & 75 & 278 & 213 & {-} & {-} & {-} & {-} & 2,384 & 374\\
   \cline{1-11}
   \multicolumn{1}{l|}{{Total images}}  & 18,045 & 191 & 354 & 305 & 134 & 3,390 & 9,500 & 1,310 & 2,384 & 467\\
   \Xhline{1.2pt}
  \end{tabular}}
\end{table*}

\section{Dataset}
\label{sec4}
Datasets used for this study can be divided into three: dataset for pre-training backbone, dataset for classification, dataset for severity quantification.

\subsection{Dataset for Pre-training}
For the pre-training of the backbone netowrk to extract the low-level CXR features, we used CheXpert dataset containing 10 labeled CXR findings: no finding, cardiomegaly, opacity, edema, consolidation, pneumonia, atelectasis, pneumothorax, pleural effusion, and support device. With a total of 224,316 CXR images from 65,240 subjects, the 32,387 lateral view images were excluded, leaving 29,420 PA and 161,427 AP view data available. With this large number of CXRs, it was able to train the backbone network robust to the variation in subjects, which is one of the key strengths of our model.

\subsection{Dataset for Classification}
Table \ref{table:dataset_PA} and Table \ref{table:dataset_AP} summarizes dataset resources and partitioning for PA and AP views, respectively. To train and evaluate the Transformer model, we utilized both public datasets containing labeled cases of infectious disease (Valencian Region Medical Image Bank [BIMCV] \citep{de2020bimcv}, Brixia \citep{signoroni2020end}, National Institutes of Health [NIH] \citep{wang2017chestx}) and deliberately collected CXR data from four hospitals (Asan Medical Center [AMC], Seoul, Korea; Chonnam National Univerity Hospital [CNUH], Daejeon, Korea; Yeungnam University Hospital [YNU], Daegu, Korea; Kyungpook National University Hospital [KNUH], Daegu, Korea) labeled by board-certified radiologists for this study. Finally, the integrated dataset was divided into three label classes including normal, other infections (e.g. bacterial infection, tuberculosis) and COVID-19 infection, considering the application in real clinical setting. For PA view images, we put 3 institutional data (CNUH, YNU, KNUH) aside as external test datasets to evaluate the generalization capability by using data collected from independent hospitals with different devices and settings. On the other hand, CNUH data was solely used as the external test dataset for AP view images, since it was the only dataset containing all three label classes.

\subsection{Dataset for Severity Quantification}
Table \ref{table:dataset_sev} summarizes dataset resources and global severity levels. Different from diagnosis, the PA and AP view data were integrated and utilized without division for severity quantification task, since there is possibility that follow-up images may be obtained with both PA and AP view even in a single patient. Two board certified radiologists labeled the severity for three institutional datasets (CNUH, YNU, KNUH) using the array-based severity labeling method of \citet{toussie2020clinical} as in Fig. \ref{fig6}. We also utilized publicly available data, Brixia dataset, after translating its severity score the same as that of the institutional datasets. We alternately used one institutional dataset as an external testset and trained the models with two remaining datasets together with the Brixia dataset to evaluate the generalization capability in various external settings. Besides, 12 COVID-19 cases from BIMCV dataset were used to compare the severity map generated by our model to those annotated by clinical experts.

\begin{table}[!h]
  \caption{Summary of dataset resource and global severity level for severity labeled CXRs.}
  \label{table:dataset_sev}
  \centering
  \resizebox{0.4\textwidth}{!}
  {\small
  \begin{tabular}{c|c|c|c|c|c}
  \Xhline{1.2pt}
  {\textbf{Severity}} & {\textbf{Total}} & {\textbf{CNUH}} & {\textbf{YNU}} & {\textbf{KNUH}} & {\textbf{Brixia}} \\
  \cline{1-6}
  \hline
   \multicolumn{1}{c|}{1}  & 361   & 26   & 63  & 25   & 247    \\
   \multicolumn{1}{c|}{2}  & 521   & 11   & 59  & 22   & 429    \\
   \multicolumn{1}{c|}{3}  & 448   & 8    & 25  & 18   & 397    \\
   \multicolumn{1}{c|}{4}  & 920   & 7    & 35  & 31   & 847    \\
   \multicolumn{1}{c|}{5}  & 774   & 12   & 18  & 29   & 715    \\
   \multicolumn{1}{c|}{6}  & 1,758 & 17   & 86  & 171  & 1,484  \\
   \cline{1-6}
   \multicolumn{1}{c|}{{Total}}  & 4,782 & 81   & 286 & 296  & 4,119 \\
   \Xhline{1.2pt}
  \end{tabular}}
\end{table}

\begin{table*}[!t]
  \caption{Diagnostic performance of the proposed model in various external test datasets from three different institutions for PA view.}
  \label{table:ext_PA}
  \centering
  \resizebox{0.9\textwidth}{!}
  {%
  \begin{tabular}{M{1.06cm}|M{1.02cm}|M{1.0cm}|M{1.04cm}|M{1.0cm}|M{1.02cm}|M{1.0cm}|M{1.04cm}|M{1.0cm}|M{1.02cm}|M{1.0cm}|M{1.04cm}|M{1.0cm}}
  \Xhline{1.2pt}
  \multicolumn{1}{l|}{\textbf{Metrics}} & \multicolumn{4}{c|}{\textbf{External dataset 1 (CNUH)}} & \multicolumn{4}{c|}{\textbf{External dataset 2 (YNU)}} & \multicolumn{4}{c}{\textbf{External dataset 3 (KNUH)}} \\
  \cline{2-13}
  & {Avg.} & {Normal} & {Others} & {COVID} & {Avg.} & {Normal} & {Others} & {COVID} & {Avg.} & {Normal} & {Others} & {COVID} \\
  \hline
   \multicolumn{1}{l|}{AUC} & 0.932 & 0.938 & 0.926 & 0.931 & 0.921 & 0.947 & 0.908 & 0.907 & 0.928 & 0.955 & 0.908 & 0.921 \\
  \multicolumn{1}{l|}{Sensitivity} & 83.4 & 84.7 & 82.1 & 83.3 & 88.4 & 92.3 & 85.4 & 87.5 & 85.4 & 87.3 & 87.7 & 81.3 \\
   \multicolumn{1}{l|}{Specificity} & 85.3 & 88.9 & 88.3 & 78.6 & 84.2 & 90.8 & 82.5 & 79.3 & 86.8 & 89.4 & 82.1 & 88.8 \\
   \multicolumn{1}{l|}{Accuracy} & 83.8 & 85.2 & 87.7 & 78.6 & 84.9 & 91.8 & 83.4 & 79.4 & 86.9 & 88.3 & 84.3 & 88.1 \\
   \Xhline{1.2pt}
  \end{tabular}}  
\end{table*}

\subsection{Details of Implementation and Evaluation}
The CXR images were preprocessed via histogram equalization, Gaussian blurring with $3 \times 3$ kernel, normalization, and finally resized to $512 \times 512$. As our backbone network, the modified version of the network proposed by \citet{ye2020weakly}, which comprises DenseNet-121 baseline followed by PCAM operations. Among several layers of intermediate feature maps, we used the feature map of size $16 \times 16 \times 1024$ just before the PCAM operation. For subsequent Transformer architecture, we used standard Transformer model with 12 layers and 12 heads per each layer, while comparison with Transformer architecture with different network size is also provided in Section \ref{sec5}.6.1. 

For pre-training of the backbone network, Adam optimizer with learning rate of 0.0001 was used. We trained the backbone network for 160,000 optimization steps with step decay scheduler with batch size of 8. For training of classification model, SGD optimizer with momentum 0.9 was used with learning rate of 0.001. A max gradient norm of 1 was applied to stabilize training. We trained the model for 10,000 optimization steps with cosine warm-up scheduler (warm-up steps = 500) with batch size of 16. We trained two individual classification models for PA and AP view images, respectively. For severity quantification, a map head with four upsizing convolutional blocks is used, with last block followed by sigmoid non-linearity which squashes output into [0-1] range. Training of severity quantification model was done with SGD optimizer with learning rate of 0.003 for 12,000 optimization steps with constant learning rate, and batch size of 4 was used. These optimal hyperparameters were determined experimentally. 



We used the area under receiver operating characteristic curve (AUC) as the evaluation metrics for diagnostic performance of the classification model, but also calculated sensitivity, specificity, and accuracy after adjusting the thresholds to meet the sensitivity value of $\geq 80 \%$, if possible. As evaluation metrics for severity quantification, we used the Mean Squared Error (MSE) as the main metric, but the Mean Absolute Error (MAE), Correlation Coefficient (CC), and $R^2$ score were also measured and compared. For classification model, the comparison between models and ablation studies were performed with PA view data, since it usually offers better diagnostic quality and is standard position for CXR diagnosis of lung disease, while both PA and AP view CXRs were used in severity quantification model. 

All experiments including preprocessing, development and evaluation of the model, was performed using Python version 3.7 and Pytorch library version 1.7 on Nvidia Tesla V100 and NVidia RTX 3090.

\section{Experimental results}
\label{sec5}

\subsection{Diagnostic Performance on External Test Datasets}
The diagnostic performances of the proposed model for PA view images are provided in Table \ref{table:ext_PA}. On average of 3 label classes (normal, other infection, COVID-19), our model showed the mean AUCs of 0.932, 0.947, 0.928, sensitivities of 83.4\%, 88.4\%, 85.4\%, specificities of 85.3\%. 84.2\%, 86.8\%, and accuracy of 83.8\%, 84.9\%, 86.9\% in three external institutions, which confirmed the stable performance (AUC $\geq 0.900$) and the generalization capability of our method in clinical situations with different devices and settings. For AP view images, our model showed mean AUCs of 0.890, 0.880, 0.828 for each 3 label classes (Table \ref{table:ext_AP}), which was slightly decreased compared to those of PA view images but still showed fair performance (AUC $\geq 0.800$) in the external test dataset, considering the fact that the diagnosis of infectious disease using only AP view image is not standard and usually deteriorates the diagnostic performance.

\begin{figure}[!t]
\centering
\includegraphics[width=0.43\textwidth]{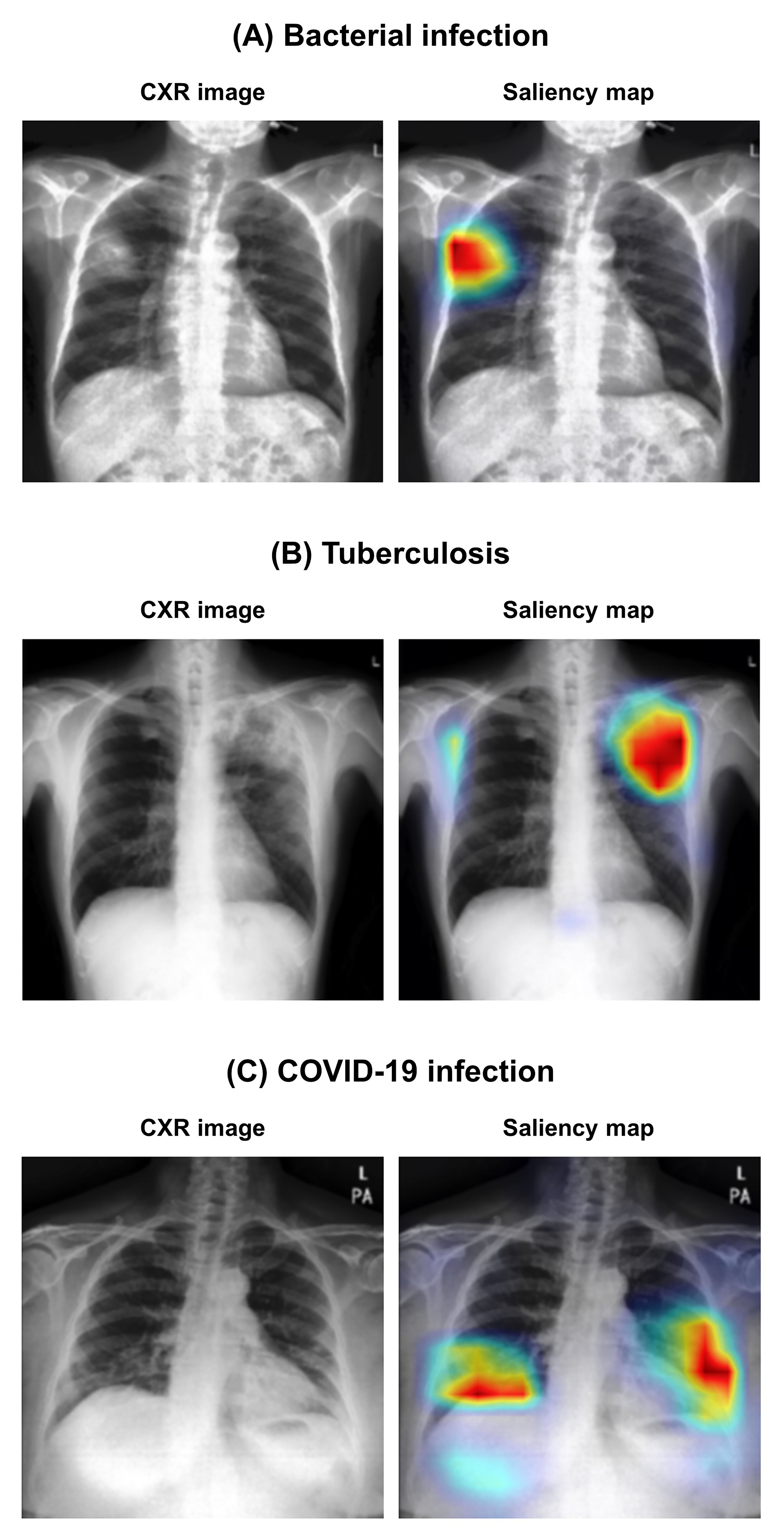}
\caption{Examples of visualization results for each disease classes. (A) Bacterial infection, (B) tuberculosis, and (C) COVID-19 infection.} 
\label{fig5}
\end{figure}


\begin{table}[!h]
  \caption{Diagnostic performance in external test dataset for AP view.}
  \label{table:ext_AP}
  \centering
  \resizebox{0.45\textwidth}{!}
  {%
  \begin{tabular}{M{1.4cm}|M{1.4cm}|M{1.4cm}|M{1.4cm}|M{1.4cm}}
  \Xhline{1.2pt}
  \multicolumn{1}{l|}{\textbf{Metrics}} & \multicolumn{4}{c}{\textbf{External dataset (CNUH)}} \\
  \cline{2-5}
  & \multicolumn{1}{c|}{Avg.} & \multicolumn{1}{c|}{Normal} & \multicolumn{1}{c|}{Others} & \multicolumn{1}{c}{COVID} \\
  \hline
   \multicolumn{1}{l|}{AUC} & 0.866 & 0.890 & 0.880 & 0.828 \\
   \Xhline{1.2pt}
  \end{tabular}}  
\end{table}

\subsection{Model Interpretability Results}
Fig. \ref{fig5} exemplifies the visualization of saliency maps for each disease classes in the external test datasets. As shown in the examples, our model well-localized a focal infected area either by bacterial infection (Fig. \ref{fig5} (a)) or tuberculosis (Fig. \ref{fig5} (b)), while it was also able to delineate the multi-focal lesions in periphery of both lower lungs in Fig. \ref{fig4} (c), which is typical findings for COVID-19 pneumonia.

\begin{figure*}[!t]
\centering
\includegraphics[width=0.90\textwidth]{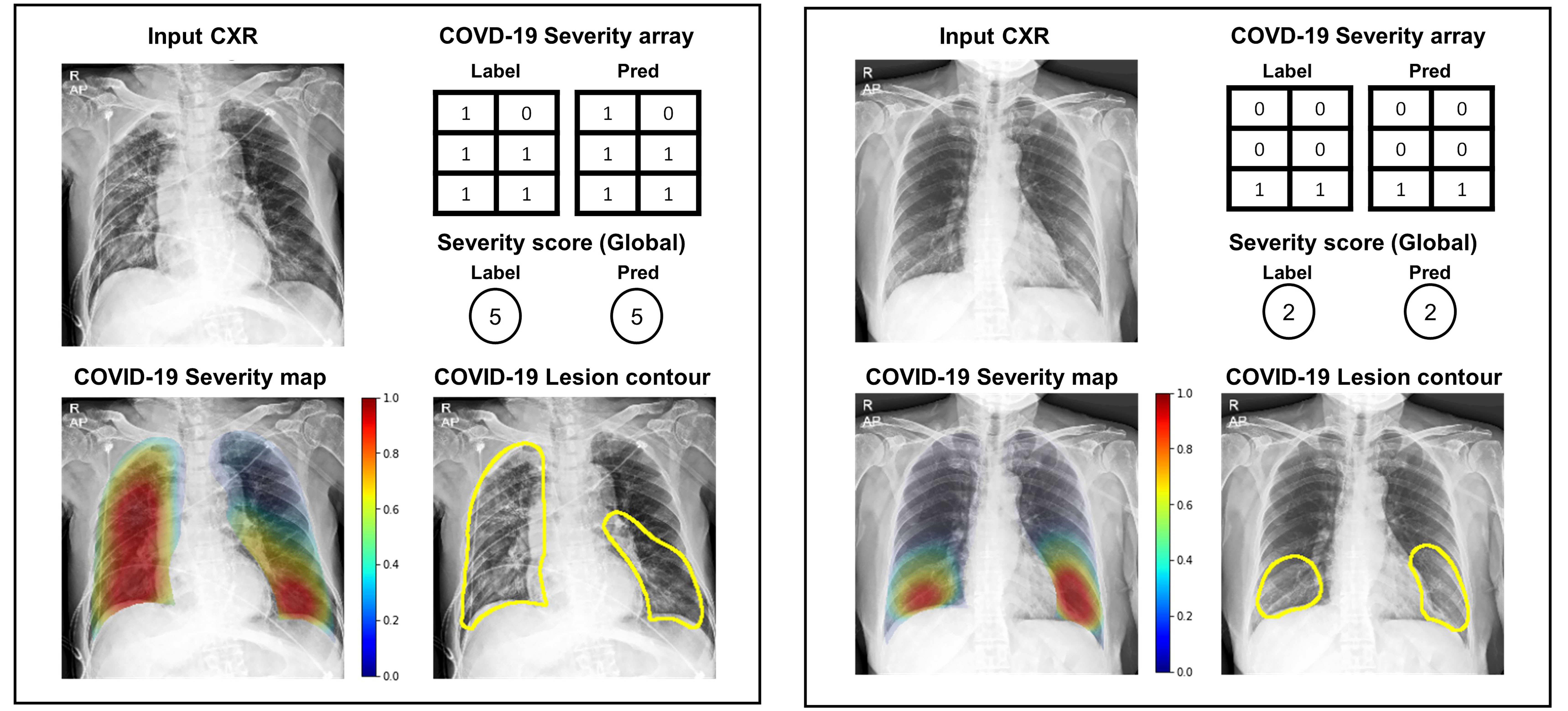}
\caption{Examples of severity quantification results of our models on CNUH external dataset.} 
\label{fig6}
\end{figure*}

\subsection{Severity Quantification Results on External Test Datasets}
The results of severity quantification of our model are shown in Table \ref{table:ext_sev}. Our model showed   the MSE of 1.682, 1.677, 1.607, the MAE of 1.028, 1.102, 0.930, correlation coefficient of 0.781, 0.777, 0.682, and $R^2$ score of 0.572, 0.572, 0.432 in three external institutions. Brixia dataset contains a consensus subset of 150 CXR images labeled by five independent radiologists. Within this subset, the average MSE between the consensus severity score calculated from majority voting and each radiologist's rating is 1.683. As a result, the MSEs of 1.657, 1.696, and 1.676 in three external institutions show our model's performance comparable to those of experienced radiologists and generalization capability in the clinical environment.

Fig \ref{fig6} illustrates the examples of severity quantification, including the predicted scores, arrays, maps, and lesion contours in one of the external test datasets, which confirms that not only can our model correctly predict global severity, but it also generates an intuitive severity map that highlights the affected area, which can also be used to contour lesions.

\begin{figure}[!t]
\centering
\includegraphics[width=0.45\textwidth]{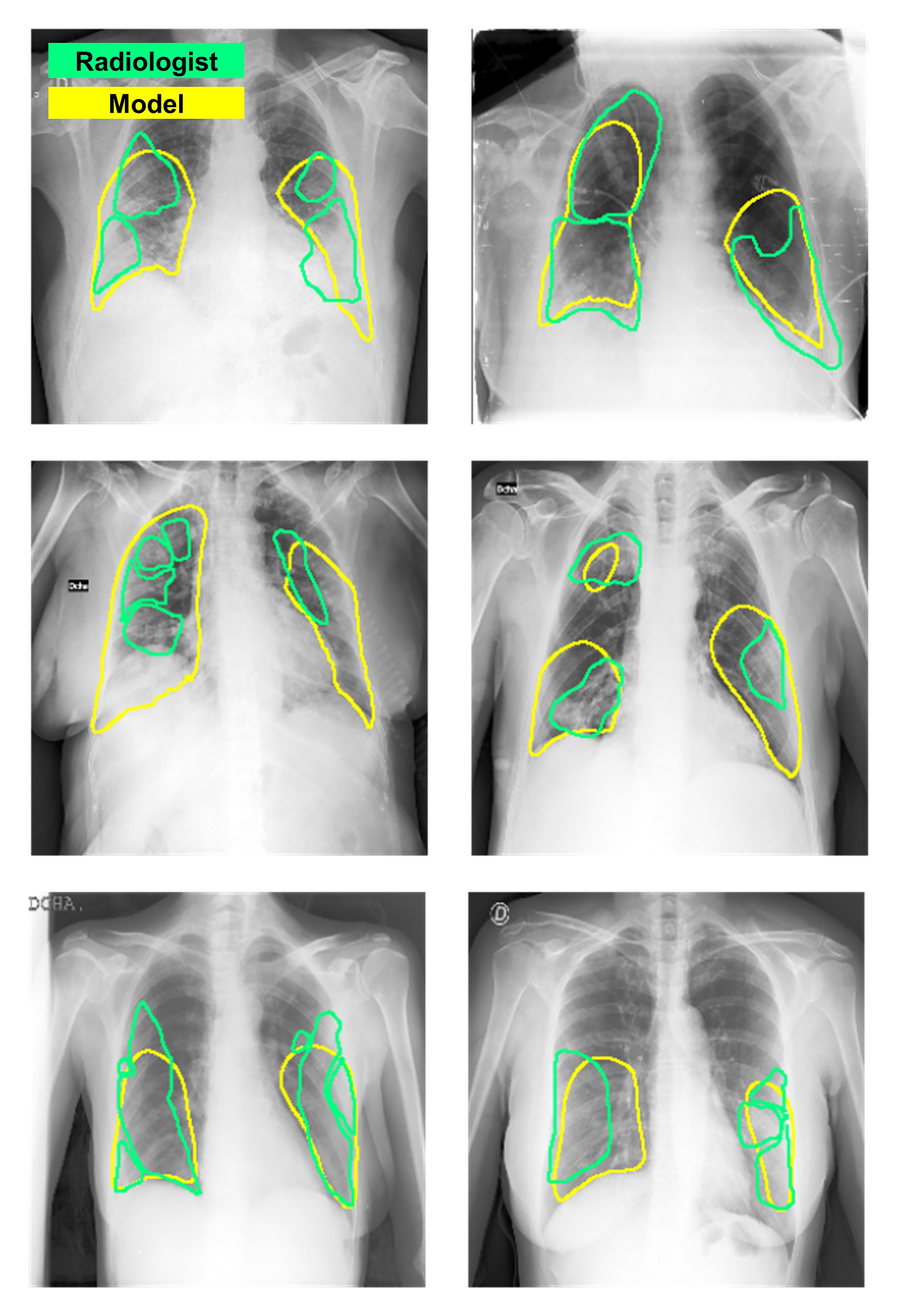}
\caption{Comparison of localization results in BIMCV dataset. Green: radiologist's annotation. Yellow: model's prediction after thresholding.} 
\label{fig7}
\end{figure}

Finally, Fig \ref{fig7} exemplifies the comparison between ground truth segmenteation label of involved area and model's prediction of involvement in BIMCV dataset. As shown in the figure, the model generally well-localized the areas of involvement.

\begin{table}[!t]
  \caption{Severity quantification performance of the proposed model in various external test datasets from different institutions.}
  \label{table:ext_sev}
  \centering
  \small
  \resizebox{0.48\textwidth}{!}
  {%
  \begin{tabular}{l|c|c|c}
  \Xhline{1.2pt}
\multicolumn{1}{l|}{\textbf{Metrics}} 
& \thead{\textbf{External dataset 1}\\\textbf{(CNUH)}} 
& \thead{\textbf{External dataset 2}\\\textbf{(YNU)}}
& \thead{\textbf{External dataset 3}\\\textbf{(KNUH)}} \\ \hline
\multicolumn{1}{l|}{MSE}   & 1.682   & 1.677   & 1.607 \\
\multicolumn{1}{l|}{MAE}   & 1.028   & 1.102   & 0.930 \\
\multicolumn{1}{l|}{CC}    & 0.781   & 0.777   & 0.682 \\
\multicolumn{1}{l|}{$R^2$}   & 0.572   & 0.572   & 0.432 \\
   \Xhline{1.2pt}
   \end{tabular}}  
\end{table}

\begin{table*}[!t]
  \caption{Comparison with CNN and other Transformer-based Models.}
  \label{table:comparison}
  \centering
  \resizebox{\textwidth}{!}
  {%
  \begin{tabular}{M{1.06cm}|M{1.02cm}|M{1cm}|M{1.04cm}|M{1cm}|M{1.02cm}|M{1cm}|M{1.04cm}|M{1cm}|M{1.02cm}|M{1cm}|M{1.04cm}|M{1cm}}
  \Xhline{1.2pt}
   \multicolumn{1}{l|}{\textbf{Methods}} & \multicolumn{4}{c|}{\textbf{External dataset 1 (CNUH)}} & \multicolumn{4}{c|}{\textbf{External dataset 2 (YNU)}} & \multicolumn{4}{c}{\textbf{External dataset 3 (KNUH)}} 
  \\
  \cline{2-13}
  & Avg. & {Normal} & {Others} & {COVID} & Avg. & {Normal} & {Others} & {COVID} &Avg. & {Normal} & {Others} & {COVID} \\
  \hline
  \multicolumn{1}{l}{\textbf{CNN-based}} \\
   \multicolumn{1}{l|}{ResNet-50} & 0.843 & 0.725 & 0.883 & 0.922 & 0.816 & 0.768 & 0.914 & 0.766 & 0.846 & 0.755 & 0.874 & 0.910 \\
  \multicolumn{1}{l|}{ResNet-152} & 0.823 & 0.755 & 0.893 & 0.821 & 0.841 & 0.811 & 0.924 & 0.789 & 0.828 & 0.795 & 0.800 & 0.890 \\
   \multicolumn{1}{l|}{DenseNet-121} & 0.852 & 0.740 & 0.907 & 0.908 & 0.880 & 0.820 & \textbf{0.937} & 0.883 & 0.835 & 0.795 & 0.792 & 0.920 \\
   \hline
   \multicolumn{1}{l}{\textbf{Transformer-based}} \\
   \multicolumn{1}{l|}{ViT} & 0.885 & 0.883 & 0.879 & 0.894 & 0.900 & 0.901 & 0.906 & 0.895 & 0.838 & 0.855 & 0.753 & 0.906 \\
   \multicolumn{1}{l|}{ViT(hybrid)} & 0.857 & 0.865 & 0.848 & 0.858 & 0.902 & 0.936 & 0.885 & 0.883 & 0.857 & 0.889 & 0.760 & \textbf{0.922} \\
   \multicolumn{1}{l|}{\textbf{Ours}} & \textbf{0.932} & \textbf{0.938} & \textbf{0.926} & \textbf{0.931} & \textbf{0.921} & \textbf{0.947} & 0.908 & \textbf{0.907} & \textbf{0.928} & \textbf{0.955} & \textbf{0.908} & 0.921 \\
   \cline{1-13}
   \Xhline{1.2pt}
  \end{tabular}} 
\end{table*}

\begin{table}[!t]
  \caption{Comparison of severity quantification performance with CNN and other Transformer based Models.}
  \label{table:comparison_sev}
  \centering
  \resizebox{0.48\textwidth}{!}
  {\large
  \begin{tabular}{l|c|c|c}
  \Xhline{1.2pt}
\multicolumn{1}{l|}{\textbf{Methods}} 
& \thead{\textbf{External dataset 1}\\\textbf{(CNUH)}} 
& \thead{\textbf{External dataset 2}\\\textbf{(YNU)}}
& \thead{\textbf{External dataset 3}\\\textbf{(KNUH)}} \\ \hline
\multicolumn{1}{l}{\textbf{CNN-based}}  & \multicolumn{1}{l}{}& \multicolumn{1}{l}{}& \multicolumn{1}{l}{} \\
\multicolumn{1}{l|}{ResNet-50}        & \textbf{1.649}   & 2.072   & 2.024  \\
\multicolumn{1}{l|}{ResNet-152}       & 1.718            & 2.135   & 2.026   \\
\multicolumn{1}{l|}{DenseNet-121}     & 1.839            & 1.963   & 1.693   \\ \hline
\multicolumn{1}{l}{\textbf{Transformer-based}}           & \multicolumn{1}{l}{}& \multicolumn{1}{l}{}& \multicolumn{1}{l}{}\\
\multicolumn{1}{l|}{ViT}              & 3.416            & 3.918   & 3.348   \\
\multicolumn{1}{l|}{ViT(hybrid)}      & 1.894            & 2.781   & 2.099   \\
\multicolumn{1}{l|}{\textbf{Ours}}    & 1.682            & \textbf{1.677}  & \textbf{1.607} \\
   \Xhline{1.2pt}
   \multicolumn{4}{l}{\footnotesize{Mean squared error (MSE) of the severity score (0-6) is chosen as the metric for comparison.} \par}
  \end{tabular}}  
\end{table}

\subsection{Comparison with CNN and Transformer-based Models}
To compare the performance with other CNN-based and Transformer-based SOTA models, ResNet-50, ResNet-152, DenseNet-121, the standard ViT and hybrid ViT models were tested with the same external test datasets. All models except ours were initialized with ImageNet pre-traiend weights as it significantly improved the performance. Other training and evaluation settings were kept the same for fair comparison. As suggested in Table \ref{table:comparison} and Table \ref{table:comparison_sev}, our model outperformed not only the CNN-base models but also the Transformer-based models in all of the external test datasets for both classification and severity quantification tasks, demonstrating superb and stable performance in real-world application. The performance improvement was not just the result of increased complexity of the model, considering that our model also considerably outperformed the model that has increased complexity (ResNet-152).

\subsection{Ablation Studies}
To get better understanding of our models, we conducted a series of ablation studies, as provided in Table \ref{table:ablation} and Table \ref{table:ablation_sev}.  More details are as follows.

\begin{table*}[!t]
  \caption{Ablation studies for classification.}
  \label{table:ablation}
  \centering
  \resizebox{1.0\textwidth}{!}
  {\large
  \begin{tabular}{l|c|c|c|c}
  \Xhline{1.2pt}
   \multicolumn{1}{l|}{\textbf{Methods}} & \multicolumn{1}{c|}{\textbf{Avg. of 3 External datasets}} & \multicolumn{1}{c|}{\textbf{External dataset 1 (CNUH)}} & \multicolumn{1}{c|}{\textbf{External dataset 2 (YNU)}} & \multicolumn{1}{c}{\textbf{External dataset 3 (KNUH)}} 
  \\
  \cline{2-5}
  & \multicolumn{1}{c|}{\textbf{Avg. AUC of 3 classes}} & \multicolumn{1}{c|}{\textbf{Avg. AUC of 3 classes}} & \multicolumn{1}{c|}{\textbf{Avg. AUC of 3 classes}} & \multicolumn{1}{c}{\textbf{Avg. AUC of 3 classes}}\\
  \hline
  \multicolumn{1}{l}{\textbf{Network size}} \\
   \multicolumn{1}{l|}{2 layers and 4 heads} & \multicolumn{1}{c}{0.921} & \multicolumn{1}{c}{0.914} & \multicolumn{1}{c}{0.921} & \multicolumn{1}{c}{0.929} \\
  \multicolumn{1}{l|}{4 layers and 8 heads} & \multicolumn{1}{c}{0.915} & \multicolumn{1}{c}{0.914} & \multicolumn{1}{c}{0.916} & \multicolumn{1}{c}{0.915} \\
   \multicolumn{1}{l|}{12 layers and 12 heads \textbf{(ours)}} & \multicolumn{1}{c}{\textbf{0.927}} & \multicolumn{1}{c}{0.932} & \multicolumn{1}{c}{0.921} & \multicolumn{1}{c}{0.928} \\
   \hline
  \multicolumn{1}{l}{\textbf{Intermediate feature map}} \\
   \multicolumn{1}{l|}{feature before PCAM \textbf{(ours)}} & \multicolumn{1}{c}{\textbf{0.927}} & \multicolumn{1}{c}{0.932} & \multicolumn{1}{c}{0.921} & \multicolumn{1}{c}{0.928} \\
   \multicolumn{1}{l|}{1 feature after PCAM} & \multicolumn{1}{c}{0.921} & \multicolumn{1}{c}{0.923} & \multicolumn{1}{c}{0.908} & \multicolumn{1}{c}{0.933} \\
   \multicolumn{1}{l|}{3 features after PCAM} & \multicolumn{1}{c}{0.903} & \multicolumn{1}{c}{0.901} & \multicolumn{1}{c}{0.897} & \multicolumn{1}{c}{0.910} \\
   \multicolumn{1}{l|}{10 features after PCAM} & \multicolumn{1}{c}{0.907} & \multicolumn{1}{c}{0.903} & \multicolumn{1}{c}{0.919} & \multicolumn{1}{c}{0.900} \\
   \hline
  \multicolumn{1}{l}{\textbf{Freezing backbone weights}} \\
  \multicolumn{1}{l|}{Trainable backbone weights \textbf{(ours)}} & \multicolumn{1}{c}{\textbf{0.927}} & \multicolumn{1}{c}{0.932} & \multicolumn{1}{c}{0.921} & \multicolumn{1}{c}{0.928} \\
   \multicolumn{1}{l|}{Frozen backbone weights} & \multicolumn{1}{c}{0.919} & \multicolumn{1}{c}{0.923} & \multicolumn{1}{c}{0.919} & \multicolumn{1}{c}{0.915} \\
   \hline
  \multicolumn{1}{l}{\textbf{Image size}} \\
  \multicolumn{1}{l|}{512 $\times$ 512 \textbf{(ours)}} & \multicolumn{1}{c}{\textbf{0.927}} & \multicolumn{1}{c}{0.932} & \multicolumn{1}{c}{0.921} & \multicolumn{1}{c}{0.928} \\
   \multicolumn{1}{l|}{1024 $\times$ 1024} & \multicolumn{1}{c}{0.844} & \multicolumn{1}{c}{0.887} & \multicolumn{1}{c}{0.835} & \multicolumn{1}{c}{0.809} \\
   \hline
  \multicolumn{1}{l}{\textbf{Self-supervised pre-training}} \\
   \multicolumn{1}{l|}{ViT w/o pretrain} & \multicolumn{1}{c}{0.875} & \multicolumn{1}{c}{0.885} & \multicolumn{1}{c}{0.900} & \multicolumn{1}{c}{0.838} \\
  \multicolumn{1}{l|}{ViT w pretrain} & \multicolumn{1}{c}{0.879} & \multicolumn{1}{c}{0.872} & \multicolumn{1}{c}{0.875} & \multicolumn{1}{c}{0.889} \\
  \multicolumn{1}{l|}{ViT(hybrid) w/o pretrain} & \multicolumn{1}{c}{0.872} & \multicolumn{1}{c}{0.857} & \multicolumn{1}{c}{0.902} & \multicolumn{1}{c}{0.857} \\
  \multicolumn{1}{l|}{ViT(hybrid) w pretrain} & \multicolumn{1}{c}{0.866} & \multicolumn{1}{c}{0.865} & \multicolumn{1}{c}{0.890} & \multicolumn{1}{c}{0.841} \\
  \multicolumn{1}{l|}{Proposed w/o pretrain \textbf{(ours)}} & \multicolumn{1}{c}{0.927} & \multicolumn{1}{c}{0.932} & \multicolumn{1}{c}{0.921} & \multicolumn{1}{c}{0.928} \\
  \multicolumn{1}{l|}{Proposed w pretrain \textbf{(ours)}} & \multicolumn{1}{c}{\textbf{0.928}} & \multicolumn{1}{c}{0.951} & \multicolumn{1}{c}{0.918} & \multicolumn{1}{c}{0.915} \\
   \Xhline{1.2pt}
  \end{tabular}} \\
\end{table*}

\subsubsection{Network Size}
Since the network size can be various according to the number of Transformer encoder layers and self-attention heads, we evaluated the effect of network size by constructing models with different size consisting of 2 layers and 4 heads per layer, 4 layers and 8 heads per layer, 12 layers and 12 heads per layer. The experimental results shows that the standard Transformer architecture with 12 layers and 12 heads per layer works the best.

\subsubsection{Intermediate Feature Map}
In our model, the intermediate feature map of backbone network was used as input feature corpus for Transformer architecture. Here we tested the various options for intermediate feature maps. To determine whether the features before or after PCAM operations are better, the common feature map before PCAM operation and weighted feature maps after PCAM operation were compared. In addition, in case of weighted feature maps after PCAM operation, there might be options whether to use all 10 weighted feature maps for each CXR findings to utilize all relevant features, or to use weighted feature maps of CXR findings related to infectious disease (opacity, consolidation, and pneumonia), or just to use only the most related one feature (pneumonia) to reduce redundancy. As shown in Table \ref{table:ablation}, we can see that the common feature before PCAM operation achieved best performance.

\begin{figure*}[!t]
\centering
\includegraphics[width=0.96\textwidth]{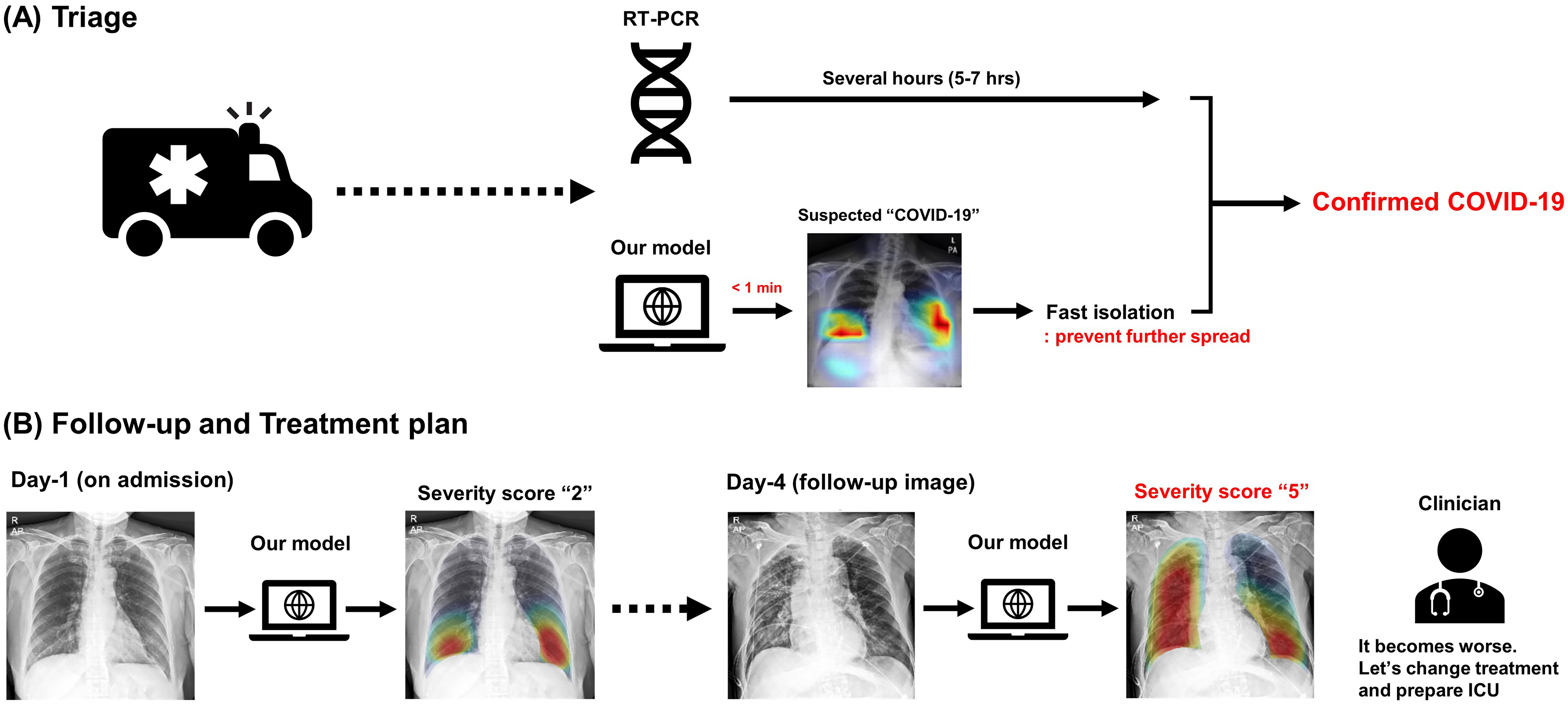}
\caption{Possible applications of our model in clinical scenario.} 
\label{fig8}
\end{figure*}

\subsubsection{Freezing Backbone Weights}
Since our model utilizes the pre-trained backbone network with large-scaled dataset, we evaluated whether freezing or leaving the backbone weights trainable yield better result. The experimental result shows that it is better to train weights than to freeze them in all three external test data sets. This could be due to an increased capacity through a trainable backbone network, which dispels the fear of overfitting.

\subsubsection{Image Size}
Regarding the input image size, we compared the input CXR of size $1024 \times 1024$ and $512 \times 512$ with the same model and experimental settings. The size of intermeidate feature map ($16 \times 16 \times 768$) remained the same by increasing the stride of patch embedding layer by two fold. As suggested in Table \ref{table:ablation}, increasing the resolution of input CXR does rather deteriorate than improve the performance, leading to severe overfitting and poor generalization performance.

\subsubsection{Self-supervised Contrastive Pre-training}
Since previous works of Transformer-based models have suggested the possibilities that self-supervised pre-training to model structural sequence may be beneficial \citep{devlin2018bert, radford2018improving, chen2020generative}, we evaluated whether Transformer-based models (standard ViT, hybrid ViT and our model) benefit from self-supervsied pre-training. We adopted SimCLR \citep{chen2020simple}, a contrastive learning technique for visual representations with data augmentation framework, as our self-supervised pre-training method. As shown in Table \ref{table:ablation}, the benefit of self-supervised pre-training was not prominent for both hybrid ViT and our models, though it slightly improved the performance of standard ViT model. Nevertheless, our model still outperformed the other Transformer-based models pre-trained with self-supervised learning, which alludes that the low-level CXR feature corpus generated by the proposed method may be more suitable input feature embedding than direct pixel-patch embedding (standard ViT) or feature embedding by ResNet backbone (hybrid ViT) for CXR classification task.

\subsubsection{ROI Pooling Type}
 ROI max-pooling method is rooted in the annotation rule where 1 or 0 is assigned to each subdivision according to the presence or absence of the abnormality, not averaging each subdivision's opacity degree. Therefore, if there are pixels with a high probability of abnormality in a small region, although most of the pixels have a low probability, the value of 1 should be assigned into the subdivision. So, the max-pooling can be more appropriate for bridging the severity map and the array than average pooling. To demonstrate the hypothesis, we compared the performance of severity quantification between the model using average-pooling and max-pooling for converting the severity map into the array. As in Table \ref{table:ablation_sev}, the ROI max-pooling method outperforms the ROI average-pooling.
 
  \begin{table}[!h]
  \caption{Ablation studies for severity quantification.}
  \label{table:ablation_sev}
  \centering
  \resizebox{0.48\textwidth}{!}
  {\Large
  \begin{tabular}{l|c|c|c|c}
  \Xhline{1.2pt}
\multicolumn{1}{l|}{\textbf{Methods}} 
& \thead{\textbf{Avg. of}\\\textbf{3 External datasets}} 
& \thead{\textbf{External dataset 1}\\\textbf{(CNUH)}} 
& \thead{\textbf{External dataset 2}\\\textbf{(YNU)}}
& \thead{\textbf{External dataset 3}\\\textbf{(KNUH)}} \\ \hline

\multicolumn{1}{l}{\textbf{ROI Pooling}} & \multicolumn{1}{l}{}& \multicolumn{1}{l}{}& \multicolumn{1}{l}{}\\
\multicolumn{1}{l|}{Avg-pool}      & \multicolumn{1}{c|}{1.846}   & \multicolumn{1}{c}{1.842}  & \multicolumn{1}{c}{1.764}    & \multicolumn{1}{c}{1.933} \\
\multicolumn{1}{l|}{Max-pool \textbf{(ours)}}   & \multicolumn{1}{c|}{\textbf{1.655}}   & \multicolumn{1}{c}{1.682}  & \multicolumn{1}{c}{1.677}    & \multicolumn{1}{c}{1.607} \\ 
\hline
\multicolumn{1}{l}{\textbf{Training data}} & \multicolumn{1}{l}{}& \multicolumn{1}{l}{}& \multicolumn{1}{l}{} \\
\multicolumn{1}{l|}{Brixia only}          & \multicolumn{1}{c|}{2.503}   & \multicolumn{1}{c}{2.392}  & \multicolumn{1}{c}{3.310}    & \multicolumn{1}{c}{1.806} \\
\multicolumn{1}{l|}{Three datasets \textbf{(ours)}}       & \multicolumn{1}{c|}{\textbf{1.655}}   & \multicolumn{1}{c}{1.682}  & \multicolumn{1}{c}{1.677}    & \multicolumn{1}{c}{1.607} \\
\Xhline{1.2pt}
\multicolumn{4}{l}{\footnotesize{Mean squared error (MSE) of the severity score (0-6) is chosen as the metric for comparison.} \par}
  \end{tabular}}  
\end{table}
 
\subsubsection{Training Data for Severity Quantification} 
As Brixia dataset constitute the majority of data for severity quantification, we evaluated whether the model can show the stable performance only with the Brixia dataset. However, as in Table \ref{table:ablation_sev}, the performance was detrimental when the institutional datasets are all excluded from training data, addressing the necessity for including at least one institutional dataset to obtain stable performance of model. This degradation in performance may be result from a variety of reasons, including the discrepancy between details in labeling method of the Brixia dataset and the institutional datasets (e.g. anatomic landmarks determining the horizontal lines), the anatomical difference between races (e.g. Caucasian vs. Asian) and so on.


\section{Discussion and Conclusion}
\label{sec6}
In this study, we developed a novel ViT model that leverages low-level CXR feature corpus for diagnosis and severity quantification of COVID-19. The novelty of this work is to decouple the overall framework into two steps, the first is to pre-train the backbone network to classify low-level CXR findings with prebuilt large-scale dataset to embed optimal feature corpus, which was then leveraged in the second step by Transformer for high-level diagnosis of disease including COVID-19. By maximally utilizing the benefit of large-scale dataset containing more than 220,000 CXR images, overfitting problem of neural network with limited numbers of COVID-19 cases can be substantially alleviated. The extensive experimental results on various external institutions have demonstrated that our model not only outperforms other CNN and Transformer-based SOTA models but also retains outstanding generalization performance irrespective of the external settings, which is sine qua non of extensive adoption of system. In addition, we also adapted the proposed method to severity quantification problem,
demonstrating a performance similar to that of clinical experts, thereby expanding its application in the clinical setting. 

In the current pandemic situation, our method holds great promise in a variety of clinical scenarios (see Fig. \ref{fig8}). Primarily, it can be used in patient triage along with the RT-PCR to isolate the suspected subjects waiting for RT-PCR results, as it was reported that positive radiological findings precede positive RT-PCR results in substantial portion (308 out of 1,014) of patients \citep{ai2020correlation}. In addition, it is possible to give guidance in treatment decision or to evaluate the response by applying our severity prediction algorithm to consecutive CXRs. Given the lack of experienced radiologists and unavailability of examination with higher sensitivity, the applications of our model could be of great value in underprivileged countries. 

Finally, the concept of making higher-level diagnosis by aggregating low-level feature corpus, which is readily available with pre-built datasets, can be applied to quickly develop a robust algorithm against newly emerging pathogen, since it is expected to share the common low-level CXR features with existing diseases.

\section{Acknowledgement}
\label{sec7}
This work was supported in part by National Research Foundation of Korea under Grant NRF-2020R1A2B5B03001980, and in part by KAIST Mobile Clinic Module Project under Grant MCM-2021-N11210024. 

\bibliographystyle{model2-names.bst}\biboptions{authoryear}
\bibliography{refs}



\end{document}